\documentclass[manuscript]{aastex}

\slugcomment{}

\shorttitle{}
\shortauthors{Abia et al.}

\begin{document}


\title{Fluorine in AGB Carbon Stars Revisited}


\author{C. Abia\altaffilmark{1}}
\affil{Departamento de F\'\i sica Te\'orica y del Cosmos, Universidad de Granada,
    18071 Granada, Spain}
\email{cabia@ugr.es}

\author{A. Recio-Blanco\altaffilmark{2} and P. de Laverny\altaffilmark{2}}
\affil{Observatoire de la C\^ote d'Azur, Dpt. Cassiop\'ee UMR 6202, 06304 Nice Cedex 4, France}

\author{S. Cristallo\altaffilmark{3}}
\affil{INAF-Osservatorio di Collurania, 64100 Teramo, Italy}

\author{I. Dom\'\i nguez\altaffilmark{1}}
\affil{Departamento de F\'\i sica Te\'orica y del Cosmos, Universidad de Granada,
    18071 Granada, Spain}

\and 

\author{O. Straniero\altaffilmark{3}}
\affil{INAF-Osservatorio di Collurania, 64100 Teramo, Italy}

\begin{abstract}
A reanalysis of the fluorine abundance in three Galactic AGB carbon stars (TX Psc, AQ Sgr and R Scl) has
been performed from the molecular HF (1-0) R9 line at 2.3358 $\mu$m. High-resolution (R$\sim 50000$) and high signal to noise  spectra obtained with
the CRIRES spectrograph and the VLT telescope or from the NOAO archive (for TX Psc) have been used. Our abundance 
analysis uses the latest generation of MARCS model atmospheres for cool carbon rich stars. Using spectral
synthesis in LTE we derive for these stars fluorine abundances that are systematically lower by
$\sim 0.8$ dex in average with respect to the sole previous estimates by Jorissen, Smith \& Lambert (1992). 
The possible
reasons of this discrepancy are explored. We conclude that the difference may rely on
the blending with C-bearing molecules (CN and C$_2$) that were not properly taken into account
in the former study. The new F abundances are in better agreement with th prediction of full network
stellar models of low mass AGB stars. These models also reproduce the $s$-process elements
distribution in the sampled stars. This result, if confirmed in a larger sample of AGB stars, 
might alleviate the current difficulty to explain the largest [F/O] ratios found by Jorissen et al. 
In particular, it may not be necessary to search for alternative nuclear chains affecting the 
production of F in AGB stars.

\end{abstract}



\section{Introduction}
The origin  of the sole stable  isotope of fluorine,  $^{19}$F, is not
well  known.  It  is easily
destroyed by proton and alpha capture reactions in the stellar interiors so that, 
its abundance is the lowest among the
light  elements  with atomic number $6\leq Z\leq  20$. Fluorine has  only a few
accessible atomic  and molecular lines suitable  for abundance studies,
detected whether in hot gaseous nebulae in the ultra-violet spectral region or
in  a rather crowed  region at  $\sim 2.3~\mu$m  in cool  objects. Three  are
the proposed sites of $^{19}$F   production:  neutrino  spallation   on   $^{20}$Ne  in
gravitational  supernovae  (SNII)  (Woosley  \& Haxton  1988),  
hydrostatic  nucleosynthesis in the He-burning core of heavily  mass-losing Wolf-Rayet (WR) 
stars (Meynet  \&  Arnould  2000),   and  hydrostatic nucleosynthesis in the He-rich  intershell
of thermally pulsing (TP) Asymptotic Giant Branch (AGB) stars  (Forestini et al.  1992). 
Up to date, the contribution of each source to the fluorine content in the
Universe is still controversial. Renda el al. (2004)  concluded that the
inclusion of all the three components is necessary to explain
the  observed fluorine Galactic  evolution as inferred  from
abundance determinations in a small  sample of field red giants (Cunha
\& Smith  2005; Cunha et al.  2008).  However,  recent measurements of
fluorine in  the interstellar medium  (Federman et al. 2005)  yield no
evidence of F  over-abundances caused by the neutrino process
in SNII.  In addition, Palacios et al.   (2005) revisited the  F production in
rotating WR star models concluding that the F yields from WR stars are
significantly  lower than previously predicted  by Meynet  \& Arnould  (2000), 
so that their contribution  to  the Galactic  F budget would be
negligible.  This leaves  AGB  stars as  the only  significant
producers.
 
The first  evidence of F production  in AGB stars was  provided by the
study  of  Jorissen,  Smith  \&  Lambert  (1992;  hereafter  JSL)  who
determined  F abundances  from rotational  HF lines  in  extrinsic (binary) and
intrinsic  stars of  near  solar metallicity  along  the AGB  spectral
sequence   (M$\rightarrow$MS$\rightarrow$S$\rightarrow$C).  JSL  found
[F/Fe] ratios  up to  $\sim 100$ times  solar and a  clear correlation
between the F enhancement and the  C/O ratio. Because the C/O ratio is
expected  to  increase  along  the  AGB evolution  due  to  the  third
dredge-up (TDU), this occurrence has been  interpreted as an evidence of the
F production in AGB stars. This result has been later supported by the
large F enhancements found in  post-AGB stars (Werner et al. 2005) and
planetary nebulae  (Zhang \& liu 2005; Otsuka et al.  2008, and references  therein), the
progeny of AGB stars. Nevertheless, up to now, nucleosynthetic models
in  AGB  stars  (e.g. Forestini  et  al.  1992;  Lugaro et  al.  2004;
Cristallo  et al.  2008)  have failed  to  quantitatively reproduce  the
largest [F/Fe,O] ratios found in the  JSL's sample. Such a discrepancy
has led to a deep revision  of the uncertainties associated with the major nuclear
reactions affecting  the production/destruction of F in  AGB stars, to a
search for alternative nuclear chains (e.g. Lugaro et al. 2004), or to
invoke non-standard mixing processes capable to increase the F production
below the convective envelope (e.g. the cool bottom process,  see 
Wasserburg, Boothroyd, \& Sackmann 1995 and Busso et al.  2007).  However, 
no satisfactory   solution  has   been  found, making this
large fluorine enhancement a major challenge for stellar nucleosynthesis.

Extant theoretical models show that the nuclear chain 
allowing the fluorine production in the He-rich and 
H-exhausted intershell zone of a TP-AGB star is 
$^{14}$N$(\alpha,\gamma)^{18}$F$(\beta^+)^{18}$O$(p,\alpha)^{15}$N$(\alpha,\gamma)^{19}$F.
Note that the $^{15}$N production production requires the existence of a few protons, 
which can eventually be produced by the $^{14}$N$(n,p)^{14}$C reaction, 
but only when neutrons are released by the $^{13}$C$(\alpha,n)^{16}$O reaction.
The latter is the main neutron source in AGB stars, giving rise to 
the $s$-process nucleosynthesis (e.g. Gallino et al. 1998; Cristallo et al. 2008). 
Therefore, F and $s$-elements are expected to correlate in TP-AGB stars undergoing TDU.
Nevertheless, some of the stars with the largest F abundances among those in the JSL sample 
are J-type carbon stars, which do not present $s$-element enhancements (Abia \& Isern
2000).

As a part of a more extended work focused on the study of the F production 
in AGB stars and its  dependence with the stellar metallicity, we present here
 new F abundance determinations in three galactic AGB
carbon stars included in the  JSL'sample. Based on the analysis of the
R9 (1-0)  HF line at $\lambda_{air}=2.33583~\mu$m\footnote{In the following, all the
wavelengths will be given in air.}, we found F  abundances on average
$\sim  0.8$ dex  lower  than those  in  JSL. We  discuss the  possible
reasons  of this discrepancy  and show  that the  new F  values nicely
agree with the state of the art of AGB nucleosynthesis models.

\section{Observations and Analysis}
Echelle spectra  in the 2.3 $\mu$m  K-band of two  well known galactic
AGB  carbon stars namely,  AQ Sgr  (N-type) and  R Scl  (J-type), were
obtained with the  8.2 m Antu telescope of ESO's  VLT on Cerro Paranal
observatory using  the CRIRES spectrograph\footnote{Observing programs
080.D-0310(A) and 081.D-0276(A),  respectively.}. The spectral range covered
was $\lambda\sim  2.290-2.450~\mu$m,  with some  gaps between the four 
detectors, including  the R9  (1-0)  line  of HF  at
$\lambda\sim  2.3358~\mu$m. This  line, in  our opinion,  is the
best one for  F abundances analysis in cool stars  as we will show
below. The selected spectral domains also contain several CO, C$_2$ and
CN lines  (and their respective  isotopic variations) that  allowed the
determination  of   the  CNO content, the C/O  and  $^{12}$C/$^{13}$C   ratios  in  the
stars. The  resolving power of the  spectra was R$\sim  50000$ and the
exposure times were chosen to achieve a S/N ratio larger than 100 at the
position  of the  R9 HF  line. The  reduction and  calibration  of the
spectra were  done with the  standard CRIRES pipeline  procedures. Hot
standard  stars at similar  air mass  were observed  immediately after
each target  object to properly  remove telluric lines using  the task
{\it telluric} within the IRAF software package. Additionally, and for comparison purposes,
we also analysed the $\sim2.3$  $\mu$m spectrum of the AGB carbon star
TX  Psc  (N-type)  and  of  the normal  (O-rich)  giant  $\alpha$  Boo
(Arcturus), both stars also included in the JSL's sample.   The spectra of
these stars were obtained from the NOAO digital public archive of high
resolution                                                     infrared
spectra\footnote{ftp://ftp.noao.edu/catalogs/hiresK/.}   and   have  a
resolving power  R$> 45000$. More details on these spectra (telescopes, dates of observation etc.) can be seen in
Wallace \& Hinkle (1996). The NOAO  spectra include many
other HF lines, among  them the R15 ($\lambda\sim 2.2826~\mu$m)
and R16  ($\lambda\sim 2.2778~\mu$m) (1-0) HF lines  used in the
JSL's  analysis of AGB  carbon stars. For  TX Psc,  we also  analysed the  optical spectrum
obtained with the  SARG spectrograph at the 3.5 m TNG  at the Roque de
los  Muchachos Observatory. The  resolving power  of this  spectrum is
$\sim 160000$ and  the S/N ratio achieved largely  exceeded 100 in the
analysed spectral  range   ($4700-8100$~{\AA}).   This allowed  us  to
derive the $s$-element abundances in this carbon star.

Atmospheric parameters were derived in the following way: for Arcturus
we adopted those deduced by  Peterson et al. (1993): T$_{eff}=4300$ K,
log g$=1.5$,  [Fe/H]$=-0.5$ and  $\xi=1.7$ kms$^{-1}$. For  the carbon
stars,  the effective  temperatures derived  from  infrared photometry
calibrations by  Bergeat et al. (2001)  were adopted. For  the other 
parameters (namely, gravity,  metallicity, microturbulence and the CNO
abundances), the  values derived by Lambert  et al. (1986) were used.  We verified,  
for instance,  that  such metallicity
values are compatible with the  metallicity that we derive from a
Fe I  line at  $\lambda\sim 2.3308~\mu$m  and from a  Na I  line at  
$\lambda\sim 2.3348~\mu$m,
both covered in  the spectral range studied here.  These stellar parameters
were considered, nevertheless, as  a starting point, the final adopted
values  (Table  1)  were  obtained  through an  iterative  process  by
comparing the observed spectra  with theoretical ones. We advance that
the stellar parameters deduced here for all the stars are very similar
to those  adopted in JSL, so  that the resulting differences  in the F
abundances  cannot be  ascribed to  that. 

We used spherically  symmetric model
atmospheres computed from a new version of the MARCS
code for cool  O-rich and C-rich stars assuming a  mass of 1 M$_\odot$
for  Arcturus, and  2  M$_\odot$ for  the  rest of  the  stars in  the
sample. Details of  this new grid of model atmospheres  can be seen in
Gustafsson et al.  (2008) and some details on the C-rich models can be found
in de Laverny et al. (2006). For the infrared spectra, the linelist used
for C$_2$ is  from Wahlin \& Plez (2005).  The  CO lines come from the  
linelist of  Goorvitch (1994), whereas  the CN and  CH lines
were assembled from the best  available data and are described in Hill
et  al. (2002)  and  Cayrel et  al.  (2004). Our  molecular list  also
include  lines of  
CaH,  SiH, FeH  and  H$_2$O taken  from the  HITRAN
database (Rothman  et al. 2005).  Unfortunately, almost no absorption  of these
molecular species is seen in  the spectra of the Sun and/or Arcturus in the
2.3 $\mu$m region,
thus they  cannot be calibrated using these  stars. Nevertheless, when
fitting the spectra of the carbon  stars in this spectral region, dominated by CO, CN and C$_2$
absorptions  (in this order),  we verified  that only  a few  of these
molecular absorptions need wavelength correction. The line list for
the HF lines is the one computed  by R.H. Tipping (unpublished) and it is
the  same  used  previously  by  JSL and  recently  by  Uthentaler  et
al. (2008).  The calculated list of  Tipping can be  regarded as quite
accurate (see the discussion in  these authors). For the optical
spectra of TX Psc, we used the same atomic and molecular line lists as
in  de  Laverny  et  al.  (2006).   We  refer  to  these  authors  for
details. Finally, the atomic lines in the infrared were taken from the
VALD  database (Kupka  et al.  2000) and  calibrated/corrected  by the
standard indirect method using the spectra of the Sun and Arcturus.

The classical  method of  spectral synthesis  in LTE  for the
abundance  analysis was used.   Theoretical  spectra  were  computed  with  the
TURBOSPECTRUM code (Alvarez \& Plez 1998, and further improvements) in
spherical geometry and convolved  with Gaussian functions to mimic the
corresponding instrumental profiles  adding a macroturbulence velocity
typically of  $\sim 5-7$  kms$^{-1}$.   A  model  atmosphere   constructed  with
preliminary parameters derived as mentioned above, was used to produce
synthetic spectra in the studied regions. Comparison of synthetic and
observed  spectra provided  new estimates  of the  CNO  abundances, in
particular of the  C/O ratio that dominates the  shape of the spectra,
and  metallicity,  which  were  then  used to  select  a  new  model
atmosphere. Several  {\it clean}  CO, CN  and C$_2$ lines are present  in the
observed spectral range allowing an independent determination of C and
O abundances and of the  C/O and $^{12}$C/$^{13}$C ratios. This region
contains also interesting  $^{12}$C$^{17}$O and $^{12}$C$^{18}$O lines
suitable  for   the  determination  of  the   oxygen  isotopic  ratios
(e.g. Harris  et al. 1987,  and below). This iterative  process, which
involved also T$_{eff}$ and gravity  (the N abundance has only a minor
role  in   the  theoretical  spectrum),   was  carried  out   until  a
satisfactory fit to the infrared  spectra (and optical, in the case of
TX Psc)  was found.  For the C-rich objects, the derived C/O ratios
in that way have an additional uncertainty due to the adopted O abundance
which we cannot determine independently. This is because theoretical spectra
are almost insensitive to a large variation in the O abundance provided that the
difference $\epsilon$(C)-$\epsilon$(O) is kept constant. Therefore, this ambiguity
allows a range of oxygen abundances and C/O ratios giving almost identical synthetic
spectra. Nevertheless, even considering an uncertainty of a factor three in the
oxygen abundance, the C/O ratios derived in our C-stars are uncertain in less than
a factor $\sim 1.2$. When the final  atmosphere parameters  were found
(see Table  1), the HF lines  were included in the  computation of the
synthetic spectrum  and we changed the F  abundance until a  good fit was
obtained to the R9  HF line (and others HF features in  the case of TX
Psc).\footnote{The sole HF line detectable in the infrared spectrum of
  $\alpha$ Boo is the R9 line. For AQ Scl and R Scl, the
  CRIRES set-up does  not allow the simultaneous covering  of other HF
  lines if the R9 is included.}

The  dependence of the  fluorine abundance  on the  stellar parameters
relies on the particular HF  line used, although it is rather similar for
all  of  them.  The results  for  the  R9  line  presented below  can  be
considered as a representative case. The changes were computed using a
baseline model for  TX Psc. A change of $+100$ K  leads to an increase
of $+0.15$ dex in the fluorine  abundance; a change of $+0.5$ in log g
produces a  variation of  $-0.15$ dex in  F and changes  of $\xi=+0.5$
kms$^{-1}$ results in a modification of $-0.15$ dex. Variations in the
metallicity of the model atmosphere  scale linearly to the F abundance
derived. This  means that  the [F/Fe] ratio  is almost  independent of
the [Fe/H] value adopted in the  model atmosphere. An interesting point is
that the R9 line is almost insensitive to moderate changes in the C/O ratio (or
the  CNO  content).  This  is  because this  line,  according  to  our
molecular and  atomic line  lists, is apparently  free of  blends (see
Figs. 1  and 2 and  the discussion below).  However, for the  other HF
lines, for instance  the R15 and R16 lines, a change  of $\pm 0.05$ dex in
the C/O ratio translates directly into a $\mp 0.05$ dex variation of the F
abundance.  This  makes  the  R9  line very  useful  for  F  abundance
derivations  in cool  C-rich objects  since in  these objects  the C/O
ratio  is the  critical  parameter determining  the  structure of  the
atmosphere and thus, the appearance  of the spectrum. The quadratic addition of
all these sources of error gives a total uncertainty of $\pm 0.30$ dex for
the absolute abundance  of F. Adding the uncertainty  of the continuum
position  and the  dispersion in  the  F abundance  when derived  from
several  lines,  a  conservative  total  error  would  be  $\pm  0.35$
dex.  However, the  abundance ratio  between F  and any  other element
certainly  is  lower  than  this   value,  since  some  of  the  above
uncertainties  cancel  out  when  deriving the  abundance  ratio:  for
instance, for the [F/Fe] ratio we estimate a total uncertainty of $\pm 0.25$
dex.  This discussion does  not include the possible systematic errors
as the uncertainty in the model atmospheres and/or N-LTE effects.

Figures  1  and 2  show  comparisons  between  observed and  synthetic
spectra for different F abundances in the spectral region of the
R9 line for the AGB carbon stars of this study. Figure 1 displays also
the comparison in the spectral ranges  of the R15 and R16 lines in the
star TX Psc.  Final abundances are summarised in Table  1 with the CNO
and  $^{19}$F values, and  the corresponding  [F/Fe] and  [F/O] ratios
derived here. The  second figure in column eight of  Table 1 shows the
 $\epsilon(^{19}$F) value  derived by  JSL in  the same  stars. We
derived also  the $^{16}$O$/^{17}$O ratio  in the carbon stars  of our
sample.   To  do   this,   we  used   the   $^{12}$C$^{17}$O  line   at
$\lambda\sim 23357$ \AA~that can be appreciated to the left wing of the HF R9
line  in Figures  1  and 2.  We derived  1240,  1200 and  4000 for  the
$^{16}$O$/^{17}$O  ratio  in the  stars  TX Psc,  AQ  Sgr  and R  Scl,
respectively. The typical uncertainty in these ratios is $\pm 500$. 
These ratios are similar  to the typical values found in
AGB C-stars by Harris et al.  (1987). For TX Psc these authors derived
1050$_{+700}^{-500}$,  which  is consistent with  the value  obtained  here.
A detailed discussion of the oxygen isotopic  ratios in these stars is outside 
the scope of the present work.
Note, nevertheless, that  the F abundance derived from  the R9 line is
not affected at all by  the actual $^{16}$O$/^{17}$O ratio.

From Figures 1 and 2 and Table  1, it is evident that the F abundances
derived here are  considerably lower that those obtained  by JSL.  The
mean difference in $\epsilon(^{19}$F) in the four stars is $-0.63$
dex (in the sense this work  minus JSL) but, excluding the O-rich star
$\alpha$ Boo,  where the F  abundances derived agree within  the error
bar, the mean  difference increase up to $-0.96$  dex. This difference
would  be somewhat  lower  ($-0.83$  dex) if  a  solar metallicity  was
adopted in  the analysis  of R  Scl. The metallicity  of this  star is
actually uncertain since we  found several combinations of the average
metallicity  and  CNO  abundances   given  a  reasonable  fit  to  its
spectrum. The theoretical fit to the observed spectrum of this star is
indeed difficult. In  Figure 2 (top) it can be  seen that the spectrum
of this star  shows broader and more asymmetric lines  as compared with the
TX Psc and AQ Sgr spectra.   Our best estimate of the metallicity in R
Scl  is [Fe/H]$=-0.5$,  but a  [Fe/H]$\sim  0.0$ value  might be  also
compatible.\footnote{We guess that a solar metallicity was used in the
  analysis of R Scl by JSL considering that for all the AGB stars in their
  sample  they  adopted the  same  stellar  parameters  than in
  Lambert et al.  (1986).} In that case, the  F abundance derived here would
increase by $\sim +0.5$ dex,  however the [F/Fe,O] ratios would remain almost
the same  as those shown in Table  1.  In the next  section we discuss
the  possible  reasons  for   the  large  discrepancy  between  the  F
abundances  derived  here  and  those  in  JSL  and  its  consequences
concerning the F production in AGB stars.

\section{Discussion}

As noted  in \S 2,  the stellar parameters  derived  here (see
Table  1) for  all the  stars  are very  similar to  those derived  by
Jorissen et  al. (1992).  The major differences  are in  the T$_{eff}$
values, which, in any case, do not exceed 150 K. This 
difference  in  T$_{eff}$  cannot  explain the difference of about  1  dex
in  the F  abundance of carbon stars.  Differences in  the  
spectroscopic parameters  of  the HF  lines
(excitation energies  and/or oscillator strength) are also discarded since we actually used
exactly the same line data than JSL. A possible  source of systematic difference might be
the  use  of  different  model  atmospheres.  JSL  used  C-rich  model
atmospheres computed  with an older  version of the MARCS  code. Actually,
the models are the same ones used in the study of the CNO content
in AGB stars by Lambert et al. (1986). We analysed our stars with this
older version  of MARCS  models and, depending  of the  actual stellar
parameters, we  found a  maximum difference of  only $\sim  -0.15$ dex
with respect to  the new generation used here  (see \S 2), in  the sense of
new models  minus the older ones.  Thus, the only  possibility left is
the existence  of significant atomic  and/or molecular blends in  the HF lines
that have been taken into acount  in a very different way in both
works.

First, it is important to  note that the large differences between our
F  abundances  and  those  by   JSL  are  found  only  in  the  C-rich
objects. For  $\alpha$ Boo, an  O-rich star, the F  abundance derived
agrees within the error  bar considering the differences in the
adopted model atmosphere parameters. This means that the cause of the
discrepancy  has to be  related with spectral features of C-bearing molecules  that become
intense when C/O$\geq 1$ in the atmosphere. This might explain
why Cunha et al. (2003), by using the R9 line, found a good agreement with JSL
in their reanalysis of some of the O-rich stars (spectral types M,  MS and S).  
Indeed, it can be seen in Figures 1 (right most panel) and 2, that in the synthetic model computed 
with no F for the R9 line, the  theoretical  pseudo-continuum almost  reaches  the relative  flux
equal to 1.  This means that, according to our atomic and molecular line list, the R9 
line is nearly free of blends, provided high resolution spectroscopy is used (R$\geq 30000$). On the contrary,
for the R15 and R16 lines (those used by JSL, see Fig. 1), the theoretical spectrum with no F
clearly shows the presence of blends.  Actually, the blending
is very important for the  R15 line (see Fig. 1 middle panel).
According to our molecular line list both, the R15 and R16 lines, are
affected by several CO, CN and  C$_2$ absorptions. In the R16 case, the most important 
contributing feature   seems  to   be   a   $^{12}$C$^{12}$C  line   at
$\lambda\sim22778.775$  \AA,   while  in   the  R15  case,   a  strong
$^{12}$C$^{14}$N line  at $\lambda\sim 22827.354$ \AA.  No blends with
atomic species seem  to  exist at the position of these lines.  On the other hand, JSL  used  the  
classical  method  of
equivalent   width  measurement  and   curve-of-growth  in their analysis. 
This  method  might be  affected by  significant  systematic
errors when using low S/N  spectra and low spectral resolution. However,
this is not the case  in the JSL's observational data. These
authors indicate  that all HF lines  were checked for  blends and that
even in their  {\it clean} lines (R15 \& R16),  they take into account
the contribution of CN  lines when computing the total equivalent
widths. Thus,  the immediate conclusion is that the  molecular line list used
here and in JSL differ significantly  in the spectral region of the HF
lines. This is unfortunate because there is no way to test the quality
of  these molecular  lists:  all the  contributing C-bearing  molecular
features at the R9, R15 and R16 spectral regions are absent in the spectra of
any standard star  with well known stellar parameters  (e.g. the Sun,
Arcturus etc)  and cannot  be indirectly calibrated.  Nevertheless, we
believe that the molecular line list used here is rather complete. This
is supported by  the small  dipersion among the F
abundance  derived from several  HF lines  in TX  Psc. Indeed,  other HF lines
in addition to R9, R15 and R16, fall in the NOAO's spectrum of
TX Psc.  We also studied the  R13, R14, R17, R18,  R20, R21, R22
and R23  HF lines in this star.  The R14, R17, R18, R20 and  R21 were
finally  discarded  because  our   synthetic  spectrum  did  not  fully
reproduced the  observed features  (indicating that probably  our line
list is  incomplete there) but,  the average F abundance  derived from
the remaining lines  (i.e. R9, R13, R15, R16, R17  and R23) was $\epsilon(^{19}$F$)=4.83$
with a dispersion of only $\pm 0.11$ dex. This dispersion is much lower than
the expected  error due to  the uncertainties in the  stellar parameters,
suggesting that the values adopted here for TX Psc
are  close  to the  real  atmosphere  and,  most importantly,  that  the
possible atomic and molecular blends in the region of these six HF lines are
properly taken into account. We can safely conclude that our line list is
accurate in these spectral regions.
A further test to illustrate that the difference between the
atomic and molecular line lists used is probably the main cause of the discrepancy with
the analysis of JSL is to use TX Psc as the reference star instead of the
Sun. The choice of this star is justified considering the good fits that we obtain to its observed 
spectrum in the 2.3 $\mu$m region (see Fig. 1). In that case,
our relative abundances would be [F/H]$_{\rm TX Psc}^{this~work}= -0.18$ and $-0.28$ dex for R Scl
\footnote{For a better comparison, the [F/H]$_{\rm TX Psc}$ ratio quoted for R Scl is that obtained
assuming [Fe/H]$=0.0$ for this star. This probably was the metallicity adopted by JSL 
(see \S 2).} and AQ Sgr, respectively. These numbers have to be compared with 
[F/H]$_{\rm TX Psc}^{JSL}=-0.15$ and $-0.07$ dex obtained from the JSL analysis of this star. It is clear 
now that the relative abundances [F/H]$_{\rm TX Psc}$ between both works agree within
the error bar. 

The existence of blends in the infrared HF lines has been also reported in 
O-rich AGB stars.
Recently, Uttenthaler et al. (2008) studied the F abundance in the bulge O-rich
star M1347 from 10 lines of the (1-0) band of HF. They found peak
differences in the F abundance from line to line up to 0.8 dex, with an average dispersion
of $\sim\pm 0.3$ dex. In this case, because the O-rich nature of this star, the main 
blending source is probably a veiling of O-bearing molecules. Finally, we have
to stress a weakness point in this reasoning. The stars for which JSL reported the largest 
F abundances (and [F/Fe,O] ratios) are carbon stars of
SC-type. The origin of these objects is not well established (see e.g. Guandalini et al. 2008) 
although it is 
generally accepted that they are transition objects between S and C stars. Their
C/O ratio is 1 within $\pm 0.01$ (or even thousandths of dex!), therefore, according to the
above argument, the HF lines should not be very much affected by C-bearing molecular
blends. However, it should be mentioned that the structure of the atmosphere can change
dramatically with a tiny variation of the C/O ratio when it is very close to the unity. Therefore, SC
stars may be the most affected by systematic differences between the model atmosphere and the
real star. A more extended study of the F abundances in SC stars is in progress.

How such large reduction of the  F abundances in AGB carbon stars would fit
in the framework of the  current nucleosynthesis models during the AGB
phase? Obviously, before extracting any definite conclusion, a similar
analysis has to  be done in a larger sample of  AGB stars (both O-rich
and C-rich) within a wider range of metallicity. However, an immediate test
would be  the comparison between  F and $s$-element abundances  in the
same object. As mentioned in \S 1, a simultaneous production of F and
$s$-elements is expected during the  AGB phase since, for both species,
neutrons coming from  the $^{13}$C$(\alpha,n)^{16}$O reaction which is
active  in  the He-shell  during  the  thermal pulse  and interpulse  phase,  are
required.  Theoretically,  the
fluorine enhancement in the envelope is mainly determined
by the occurrence of TDU episodes, whose efficiency and number depend on the
initial metallicity of the star (the lower the metallicity is, the
deeper the TDU), on the initial mass and on the mass loss rate.
Therefore, the largest F enhancements are expected at low metallicities.
Indeed, the  F enhancements  so far found  in planetary
nebulae of different metallicities (Zhang \& Liu 2005) and in the carbon
enhanced metal poor star HE 1305+0132 (Schuler et al. 2007) agree with
this theoretical expectation. In Figure 3 we show the relative abundances with respect to
Ba  derived  in  TX  Psc. They are compared  with  recent
theoretical  nucleosythetic predictions for a  2  M$_\odot$, $Z=0.006$
TP-AGB model by Cristallo et al. (2008). The $s$-element abundances in
this star were derived from our SARG optical spectrum (see \S 2) taken
for other purposes. Details  on the method, approximations
and uncertainties of the chemical analysis can be found in our previous
works on AGB  C-stars (e.g. Abia et al. 2002; de  Laverny et al. 2006)
and  will not  be repeated  here.  We compare  observed and  predicted
abundances with respect to Ba rather than absolute enhancements because the
relative  abundances between  elements are  nearly independent  of the
details  of the  AGB modelling (such as the assumed mass loss prescription),  
and of  any possible  dilution of the stellar envelope. From Figure 3 it is evident that the observed and predicted
[X/Ba] ratios  are in a remarkable agreement when  the F abundance
derived  here in  TX Psc  is  used instead  of the  JSL's value  (open
circle). According to Cristallo et al., a ratio [F/Ba]$\sim +0.2$ (the
JSL's value) is obtained for a  2 M$_\odot$ model but at lower metallicity, $Z\sim 0.001$ (or
[Fe/H]$\sim   -1.14$).  However,   we  verified   that   the assumption of such
metallicity for TX Psc, does not allow a correct fit of the observed spectrum and, in
addition,   the  derived   relative  abundances  between   the  light
(ls) $s$-elements   (Sr,Y  and   Zr)   and  Ba are   not  reproduced   at
all. Unfortunately,  there is no  information in the  literature about
the $s$-element  content in AQ Sgr  and R Scl  but interesting enough,
the ratio [F/Fe]$=+0.09$ derived here  in R Scl is compatible with the
fact that this star is classified as a J-type carbon star, which typically do not
show  $s$-element enhancements (see  Abia \&  Isern 2000)\footnote{The
nature  of  J-type  stars   is  also  unknown  (e.g.  Lorentz-Martin
1986). They show larger  Li abundances and lower $^{12}$C/$^{13}$C($<
15$)  ratios than  the  normal N-type  carbon  stars. These  chemical
peculiarities  might be  explained  if an  extra-mixing and  burning
mechanism is operating below  the convective envelope during the AGB
phase  (the  so called  cool  bottom  process).  However, up to date
attempts to simultaneously produce  fluorine in the framework of this
mechanism have failed (e.g. Busso et al. 2007).}. Nevertheless, we can
compare  the  theoretical  [F/hs]  ratios predicted  by  the
Cristallo et al. low mass AGB models with the observed ratios for some of the MS, S
and C-stars  in the JSL's sample  (i.e. for the stars  that show some
$s$-element enhancement).  The $s$-element abundances  for these stars
can be  taken from Abia et  al. (2002) and Abia  \& Wallerstein (1998)
for  the  C-rich  objects,  and  from  Smith  \&  Lambert  (1990,  and
references therein)  for the O-rich ones. From the [Rb/ls] ratios derived in
these stars, Lambert et al. (1995) and Abia et al. (2001) conclude that most of them
are low mass ($< 3$ M$_\odot$) AGB stars. Therefore, we can safely compare
with the predictions for the 2 M$_\odot$ case. Then, 
the  observed average [F/hs] (where hs  means the average  of Ba, La,  Nd and Sm  enhancements) is
$-0.22\pm  0.26$ dex, and  $+0.46\pm 0.42$ dex for the  O-rich and  C-rich AGB
stars  in JSL,  respectively.  The  predicted [F/hs]  ratios in  the 2
M$_\odot$ $Z=Z_{\odot}$ model are $\sim -0.3$ after 4-5 TDU episodes,  when C/O$\sim 0.5-0.6$, and [F/hs]$\sim -0.6$ at
the end of the AGB (when C/O$=1.8$). That is, models and 
observations of O-rich stars are in quite good agreement, whereas   for the
carbon  stars,   there  is   almost   1  dex of discrepancy.
This discrepancy is almost cancelled when
using our new fluorine determinations (see \S2).

\section{Conclusions}
Fluorine abundances  are derived in  three Galactic AGB carbon stars  of near
solar metallicity using the R9 (1-0) HF molecular line and the state
of the art of model atmospheres for cool C-rich objects. We show that this
line is not very much affected by atomic and molecular blends and
therefore is probably the best line to derive F abundances in cool
carbon-rich objects. Our results show that the fluorine  abundance is 
systematically reduced in AGB carbon stars by $\sim 0.8$ dex with respect 
to the sole previous analysis in this kind of stars by Jorissen, Smith \& Lambert (1992). For the star TX Psc, 
where we find a significant F enhancement, the new F abundance agrees   
nicely with recent nucleosynthesis models of low mass AGB stars 
of near solar metallicity. We discuss the reasons of the discrepancy with 
previous measurements and we conclude
that  most probably  blends with C-bearing molecular lines were not 
properly taken into account in the analysis by JSL. This
systematic  reduction of  the F  abundance in AGB carbon stars, if
confirmed in a  larger sample of stars, could  reduce the present
difficulty to  explain the largest  [F/Fe,O] ratios previously found
in these  stars. In  particular, it would  not be
necessary to  search for other  nuclear chains and/or  non standard
mixing mechanism  affecting the  production/destruction of F  in AGB
stars. Our results however, do  not discard AGB stars as significant F
producers, the only way to determine the real contribution of these
stars to the  F content in the Universe  being the interplay between
accurate observational data and theoretical modelling.

\acknowledgments
Part of this work was supported by the Spanish grant AYA2005-08013-C03-03 from the MEC. 
P. de Laverny and A. Recio-Blanco acknowledge the financial support of Programme 
National de Physique Stellaire (PNPS) of CNRS/INSU, France. O. Straniero and
S. Cristallo have been supported by the PRIN-MUR program 2006.
We are thankful to B. Plez for providing us molecular line lists in the observed 
infrared domain.

{\it Facilities:} \facility{ESO VLT (CRIRES)}, \facility{ORM TNG (SARG)}.

\clearpage



\begin{figure}
\epsscale{.80}
\plotone{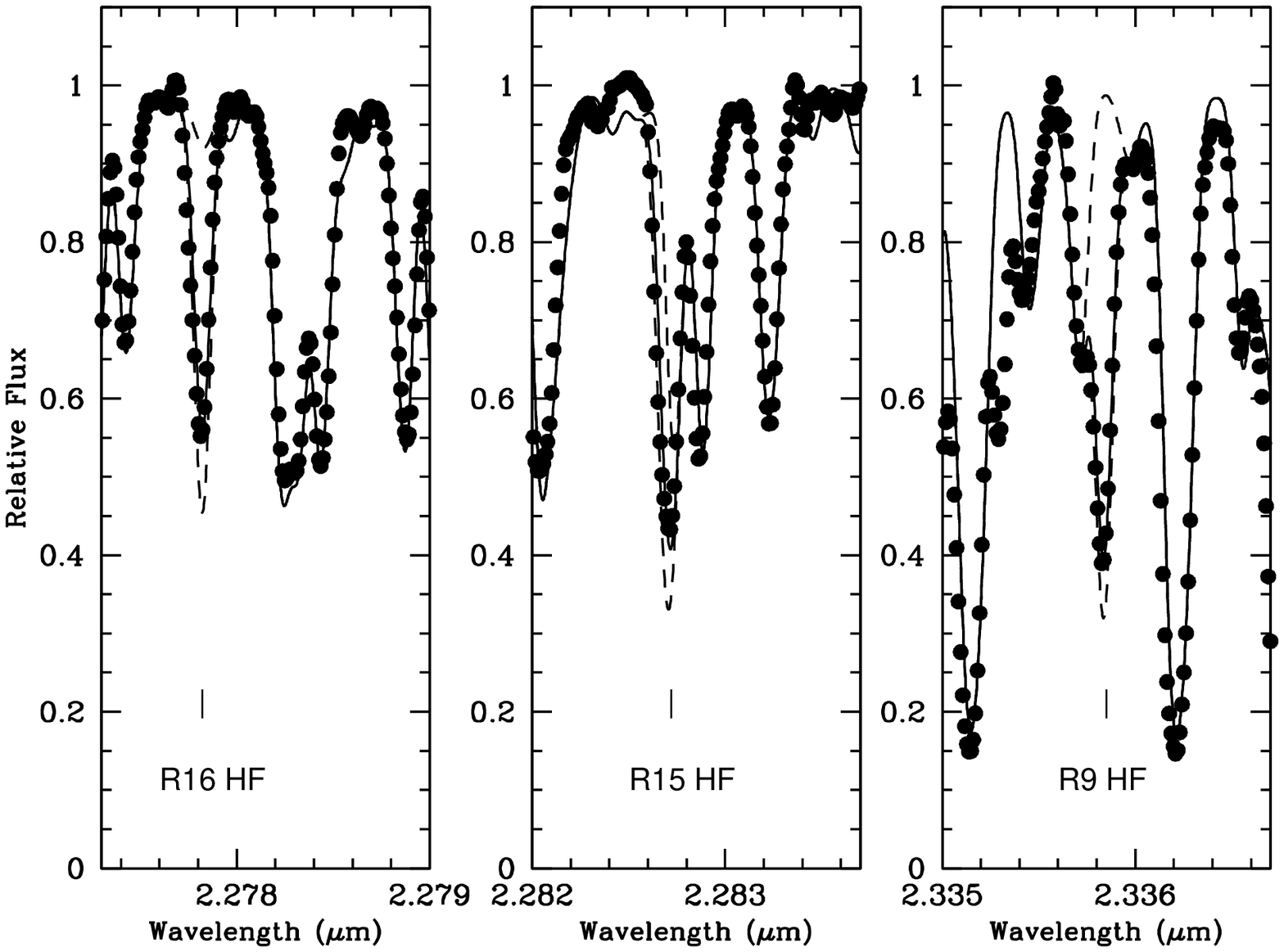}
\caption{From left to right, observed (black dots) and synthetic spectra for
  the star  TX Psc  in the  regions of the  R16, R15  and R9  (1-0) HF
  lines. In each panel, the solid line is the best fit to the HF line:
  $\epsilon(^{19}$F$)=4.9,  4.7$ and  4.9,  respectively. The  dashed  lines are  synthetic
  spectra computed with {\it no} F, and assuming the abundance derived
  by JSL, namely 5.55, using the R15 and R16 lines. The discrepancy is
  evident. The rest of the features in  the spectral ranges  shown are CO
  and CN lines. In fact the feature in the left wing of the R9 line
  (right panel) is a $^{12}$C$^{17}$O line from which we estimate a
  ratio $^{16}$O$/^{17}$O $=1240$ in this star.}
\end{figure}
\clearpage

\begin{figure}
\plotone{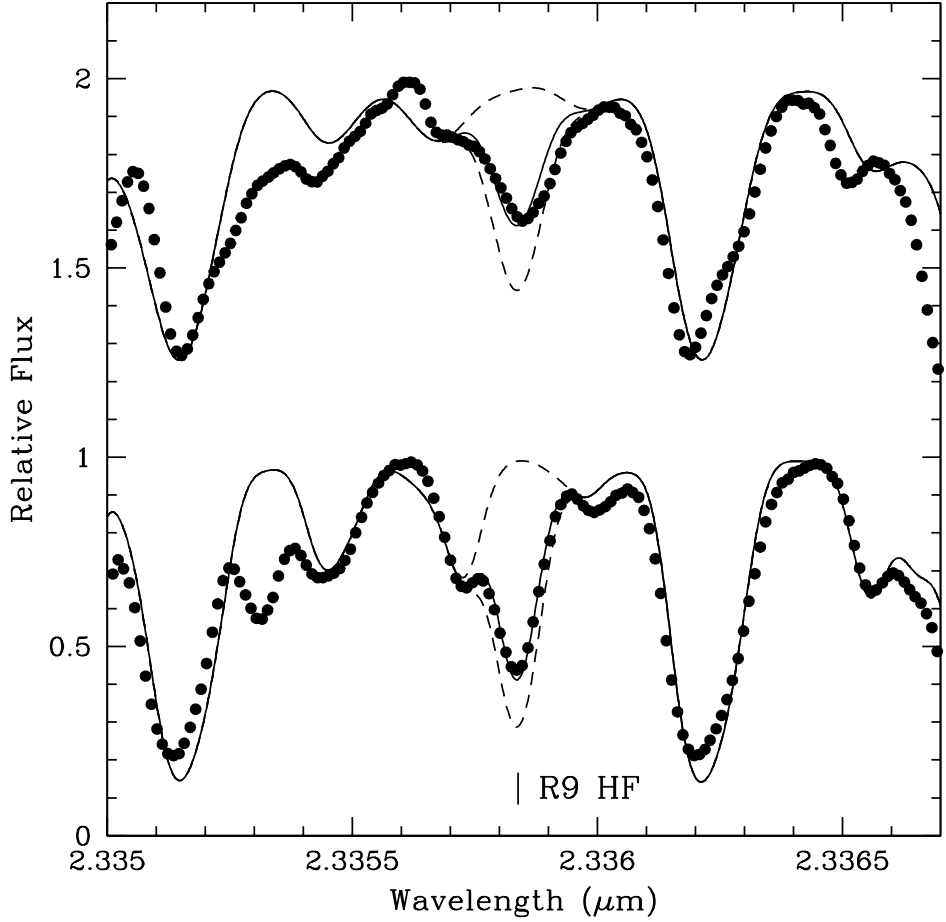}
\caption{Detail of the observed (black dots) and  synthetic spectra for the stars R
  Scl  (top) and AQ  Sgr (bottom)  in the  region of  the R9  (1-0) HF
  line. Similar to Fig. 1, solid line is the best fit: $\epsilon(^{19}$F$)=4.15$
  and  4.55,  respectively, whereas  the  dashed  lines are  synthetic
  spectra computed with {\it no} F, and assuming the abundance derived
  by  JSL,  5.40 and  5.48,  respectively.  Again  the
  discrepancy between both analysis is evident. For  these stars we derive a ratio
  $^{16}$O$/^{17}$O of 4000 and 1200, respectively. Note the arbitrary
  scale in the y-axis and the much broader lines in the 
spectrum of R Scl.}
\end{figure}

\clearpage
\begin{figure}
\plotone{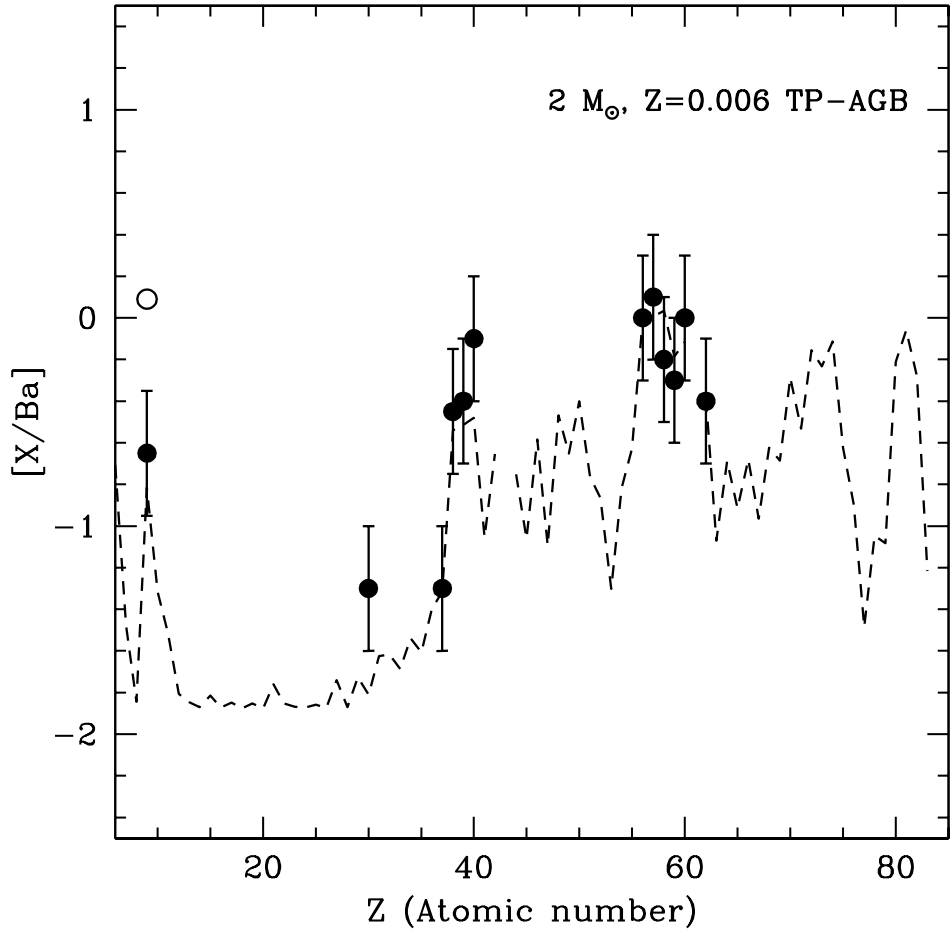}
\caption{Detailed reproduction  of the derived  $s$-element abundances
  (solid circles) in  the AGB carbon star TX  Psc with the $s$-process
  nucleosynthetic  predictions   in  a  2   M$_\odot$,  Z$=0.006$  (or
  [Fe/H]$\sim -0.36$), TP-AGB model  from Cristallo et al. (2008). All
  the abundance values  are referred to Ba.  The  open circle at Z$=9$
  is the  [F/Ba] ratio obtained  if the fluorine abundance  derived in
  JLS is  adopted. Note the much  better agreement of  the [F/Ba] ratio with
  theoretical predictions  when the F  abundance derived here  is used
  (see text for details).}
\end{figure}

\clearpage

\begin{deluxetable}{cccccccccccc}
\tabletypesize{\scriptsize}      \rotate      \tablecaption{Atmosphere
  parameters  and  fluorine  abundances} \tablewidth{0pt}  \tablehead{
  \colhead{Star}  &  \colhead{T$_{eff}$(K)}   &  \colhead{log  g  }  &
  \colhead{$\xi$ (kms$^{-1}$)} &  \colhead{[Fe/H]} & \colhead{C/N/O} & \colhead{C/O} &
  \colhead{$^{12}$C/$^{13}$C}              &  \colhead{$^{16}$O$/^{17}$O}   &      \colhead{
    $\epsilon(^{19}$F)\tablenotemark{a}}    &    \colhead{[F/Fe]}    &
  \colhead{[F/O]}  } 
\startdata  
$\alpha$ Boo  & 4300  & 1.5  &  1.7 &$-$0.5 & 8.06/7.85/8.83& 0.17  & 7 &  \nodata & 4.15, (4.01) & $+$0.10 & $-$0.07\\ 
AQ Sgr & 2800 & 0.0 & 2.3 & 0.0 & 8.79/7.76/8.78& 1.02   &  50 & 1200 &4.55, (5.48) & $+$0.10& $+$0.08\\ 
R Scl  & 2500 & 0.0 &  2.2 & $-$0.5 & 8.34/6.90/8.32& 1.05  & 20 & 4000 & 4.15, (5.40) & $+$0.09  & $-$0.07 \\ 
TX Psc & 3100  & 0.0 & 2.2 &$-$0.4& 8.83/7.72/8.82& 1.03  &  42 & 1240 & 4.83, (5.55) & $+$0.65 & $+$0.10 \\ 
\enddata
\tablecomments{The abundances in the Table are given in the usual scale $\epsilon(X)=$ log$N(X)/N(H) +12$. The solar
abundances adopted here are from Asplund et al. (2005). In particular, according to these authors, the solar fluorine abundance is $4.56\pm 0.30$.}
\tablenotetext{a}{In this column the first number is the abundance of $^{19}$F derived here (with a typical
uncertainty of $\pm 0.35$ dex), whereas the second one (in parenthesis) comes
from the analysis by JSL. The [F/Fe,O] ratios shown in the Table are obtained using the F abundances derived here.}
\end{deluxetable}




\clearpage




\end{document}